\documentclass[11pt,linenumbers,a4paper]{article}
\usepackage{jheppub}

\usepackage[usenames,dvipsnames]{color}
\definecolor{darkblue}{RGB}{0,0,196}
\usepackage[colorlinks=true]{hyperref}
\hypersetup{
    colorlinks   = true, 
    urlcolor     = blue, 
    linkcolor    = blue, 
    citecolor   = blue 
  }
\usepackage{setspace}
\usepackage{xcolor}
\usepackage{footmisc}
\usepackage[makeroom]{cancel}
\usepackage{comment}
\usepackage{lineno}
\def\be{\begin{equation}}
\def\ee{\end{equation}}
\def\ba{\begin{eqnarray}}
\def\ea{\end{eqnarray}}
\usepackage{graphicx}
\usepackage{subfigure}
\usepackage{amsmath,bbm}
\usepackage{amssymb,bm}
\usepackage{yfonts}

\usepackage{float}
\usepackage{multirow}
\usepackage{wrapfig}

\usepackage{orcidlink}

\newcommand{\dd}{ {\mathrm d} } 

\newcommand{\orcidA}{\orcidlink{0000-0003-2849-0120}} 
\newcommand{\orcidB}{\orcidlink{0000-0001-9223-6480}} 

\title{\boldmath Studying Hadronization by Machine Learning Techniques
}

\author[a,b,1]{Gábor Bíró\orcidA{}}
\author[a,c,2]{Bence Tankó-Bartalis}
\author[a,3]{Gergely Gábor Barnaföldi\orcidB{}}

\affiliation[a]{Wigner Research Center for Physics, 29--33 Konkoly--Thege Mikl\'os Str., H-1121 Budapest, Hungary.}
\affiliation[b]{Institute of Physics, E\"otv\"os Lor\'and University, 1/A P\'azm\'any P\'eter S\'et\'any, H-1117  Budapest, Hungary.}
\affiliation[c]{University of Oxford, University Offices, Wellington Square, Oxford, OX1 2JD, United Kingdom.}

\emailAdd{biro.gabor@wigner.hu}
\emailAdd{tanko.bartalis.bence@wigner.hu}
\emailAdd{barnafoldi.gergely@wigner.hu}

\abstract{
  Hadronization is a non-perturbative process, which theoretical description can not be deduced from first principles. Modeling hadron formation requires several assumptions and various phenomenological approaches. Utilizing state-of-the-art Computer Vision and Deep Learning algorithms, it is eventually possible to train neural networks to learn non-linear and non-perturbative features of the physical  processes. In this study, results of two ResNet networks are presented by investigating global and kinematical quantities, indeed jet- and event-shape variables. The widely used Lund string fragmentation model is applied as a baseline in $\sqrt{s}= 7$~TeV proton-proton collisions to predict the most relevant observables at further LHC energies.
}

\begin{document} 

\maketitle

\flushbottom


\section{Introduction}
\label{sec:introduction}

Color confinement is one of the most intriguing aspects of Quantum Chromodynamics (QCD). Due to the running nature of the strong coupling, the cross section of the partonic level scatterings can be well calculated with perturbation techniques at high energy scales. However, hadronization -- the confinement of partons into hadrons -- occurs at lower energies, where calculations can not be performed perturbatively from first principles. Investigating this non-perturbative regime requires phenomenological models, where QCD-related scaling is well hided due to the high non-linearity~\cite{Bjorken:1968dy}. 

Machine Learning algorithms are able to describe non-perturbative (non-linear) processes in high-energy physics~\cite{Feickert:2021ajf}. This raises the question whether such ML methods could provide new solutions in these soft regions, where traditionally complex Monte Carlo algorithms are used to describe hadron production. It would be also interesting to identify scaling patterns as well. Therefore, the main goal of this study is the investigation of hadronization, by applying popular Computer Vision (CV) algorithms. The primary physical motivation here is to develop a novel model of hadronization, in parallel extracting further correlations via the evaluation of Deep Neural Network (DNN) properties. 

An application-oriented feature of this investigation can have a further impact on the future data analysis. During the Long Shutdown 2 (LS2) of the Large Hadron Collider (LHC), all of the large detectors went through a major upgrade. Now, they are in the preparation phase for the Run 3, starting early 2022~\cite{Evans:2008zzb}. By approaching the High Luminosity era, it is expected to record an enormous amount of raw data, more than 200 PB each year. This is almost the quantity that was collected during Run~1 and~2 altogether. Additionally, the amount of the simulated data has to be increased along with the real experimental data, which is already a challenging task: even for a state-of-the-art hardware, to simulate only one second of LHC data, the necessary computing time is larger with a several orders of magnitude. As a solution for this ever-growing difficulty, one can already aim towards several directions:
\begin{enumerate}
  \item Development of future-generation algorithms and simulation softwares with built-in parallelization and/or hardware optimization~\cite{Grindhammer:1989zg, Biro:2019ijx, Amadio:2020ink, Butter:2020abv};
  \item Utilization of novel Machine Learning approaches~\cite{Monk:2018zsb, Vallecorsa_2018, Butter:2020abv}.
\end{enumerate}

The structure of this paper is the following: in Section~\ref{sec:jets}, the physics of hadronization and the observable quantities are summarized. The used neural network architectures are introduced in Section~\ref{sec:architecture}. The details of the training data is described in Section~\ref{sec:trainingdata}. Finally, the results on validation and prediction of global and kinematical observables, indeed jet-shape variables and event-activity classifiers are given in Section~\ref{sec:results}.

\section{Jets and event activity classifiers}
\label{sec:jets}

Monte Carlo event generators provide an essential tool for theoretical studies, phenomenological approaches, detector validations and for the planning of future facilities as well. They can combine several models together, and therefore it is possible to test the various aspects of high-energy collisions. Since it is not possible to provide a theoretical description of hadronization with the traditional perturbative approaches, it is necessary to make physically motivated assumptions that can be tuned with experimental input. Finally these can be verified with MC generators in a natural way by comparing with data again. 

There are a variety of hadronization models in the literature -- the Lund string fragmentation is one of the most successful existing model applied in the widely used {\sc Pythia} general purpose event generator~\cite{Sjostrand:1982fn, Sjostrand:2014zea}. {\sc Pythia} is able to reproduce various experimentally measured data with a high precision -- however, the true nature of hadronization itself still remains an actively studied mystery.

The experimental input for the hadronization studies are the hadronic distributions, measured by the detectors. In this study, the following quantities are considered:
\begin{enumerate}
  \item Mid-pseudorapidity density, $\dd N_{ch}/\dd\eta$
  \item Total number of (charged) particles, $N_{|y|<\pi}$ and $N_{ch, |y|<\pi}$
  \item Jet observables (with jet radius $R=0.6$ and minimum $p_{T, jet}>40$~GeV/$c$): 
  \begin{itemize}
    \item Jet transverse momentum, $p_{T, jet}$
    \item Jet invariant mass, $M_{J}^2=\left(\sum\limits_{i\in\textrm{jet}}k_i\right)^2$
    \item Jet multiplicity, $N_{J}$
    \item Jet width or radial energy profile, $\rho_{J}=\frac{\sum\limits_i\Delta R(j, p^i)p_T^i}{\sum\limits_i p_T^i}$, where $p_T^i$ is the transverse momentum of the jet constituent $i$, and $\Delta R(j, p^i)$ is the distance of the constituent $i$ and the jet axis in the $\eta, \phi$ space 
  \end{itemize}
  \item Event shape variables: the sphericity and the transverse sphericity, defined by the following equations, respectively:
  \begin{align}
    S=& \frac{3}{2}(\lambda_2+\lambda_3), \\
    S_\perp=& \frac{2\lambda_2}{\lambda_1+\lambda_2},
  \end{align}  
  where the $\lambda_i$ are the eigenvalues of the momentum tensor in Eq. (\ref{eq:momtens}):
  \begin{align}
    M_{xyz}=\sum_i \begin{pmatrix}
      p_{xi}^2 & p_{xi}p_{yi} & p_{xi}p_{zi} \\
      p_{yi}p_{xi} & p_{yi}^2 & p_{yi}p_{zi} \\
      p_{zi}p_{xi} & p_{zi}p_{yi} & p_{zi}^2
    \end{pmatrix}
    \label{eq:momtens}
  \end{align}
  and ordered as
  \begin{equation}
    \lambda_1>\lambda_2>\lambda_3,\qquad \sum_i\lambda_i=1.
  \end{equation}
\end{enumerate}
  
These observable quantities can provide detailed information about the event activity, event characteristics, and \textit{jettiness} or isotropy in an event-by-event basis. Experimentally observable, hadronic jets are important objects that carry information about the QCD hard scattering processes, occurring before the hadronization. With subtle jet analysis methods, one can extract the details of the underlying physics, such as the partonic energy loss in heavy-ion physics~\cite{Gyulassy:1990ye, Wang:2002ri}. It is, therefore, essential to measure the jet-related quantities with high accuracy.

\section{Neural network design}
\label{sec:architecture}

In traditional Computer Vision approaches, the input data is a set of digital photographs or video streams, while the output can be  e.g. the list of recognized objects and their coordinates, depending on the specific application. The most significant features of an image are typically extracted with a series of convolutional layers. 

In case of (especially at the very) deep neural networks, there is a common difficulty that occurs during the backpropagation of the learning phase, called \textit{vanishing gradient problem}~\cite{279181}. The source of the problem is that the partial derivatives of the loss function can be vanishingly small, therefore the layers at the beginning of the NN receive vanishingly small updates, resulting in a very slow train. 
There are various possible solutions for the vanishing gradient problem, including long short-term memory techniques, rectifiers as activation functions or residual networks~\cite{10.1162/neco.1997.9.8.1735}. 

In this study, a neural network based on the state-of-the-art ResNet architecture is used~\cite{he2015deep}. In these networks, an identity mapping between two distinct layers ensures that the impact of the vanishing gradients is reduced by effectively simplifying the network. In this way, it is possible to achieve higher complexity without compromising the learning speed, even if the network goes deeper. 

\begin{figure}[H]
\centering
\includegraphics[width=0.92\textwidth]{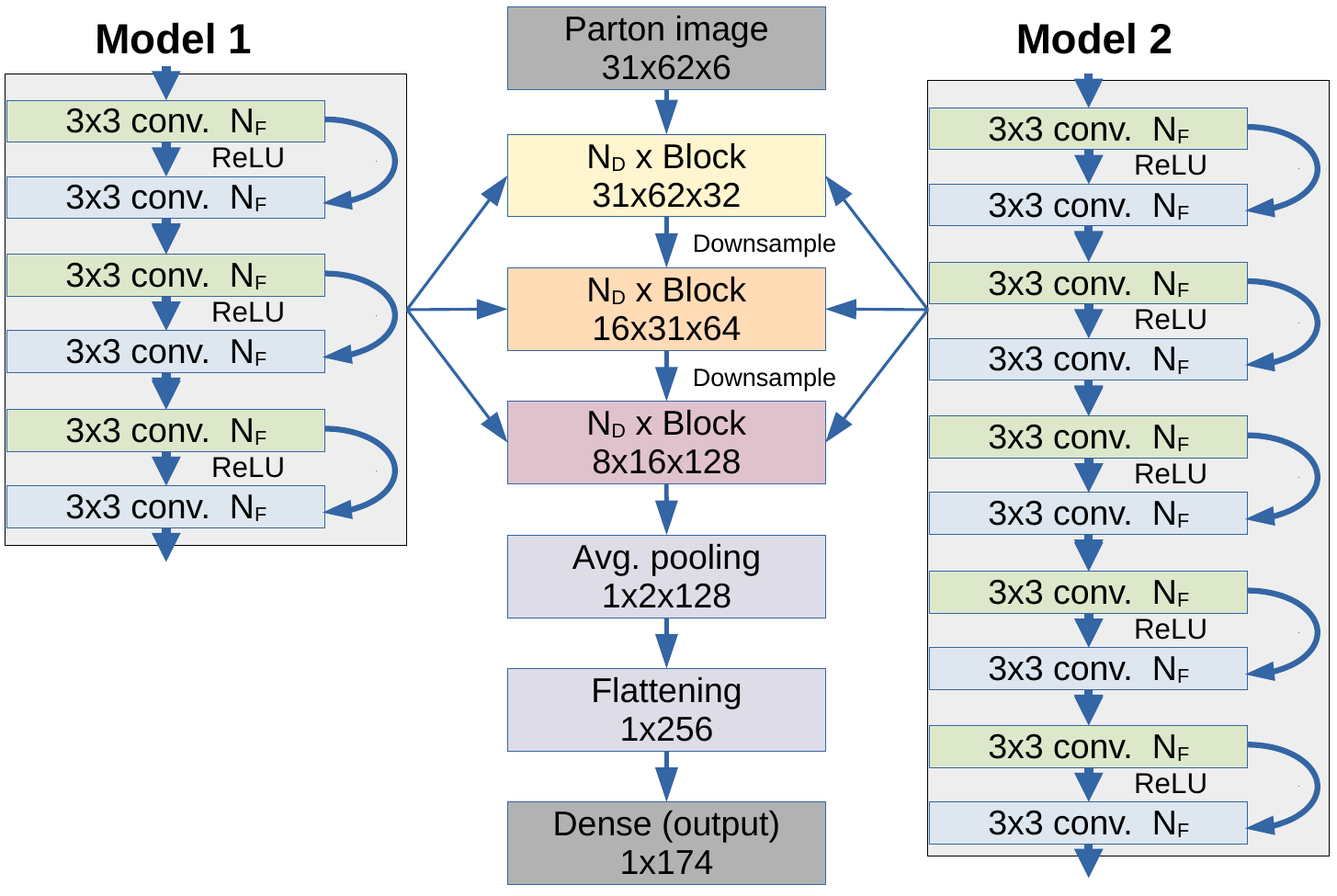}
\caption{The structure of the applied ResNet models: Model~1 (left) and Model~2 (right).}
\label{fig:layout}%
\end{figure}

The structure of the applied networks are outlined in Fig. \ref{fig:layout}. A basic building element is the \textit{residual block} with the identity mapping, with $N_F$ trainable filters of size $3\times3$, while the extent of complexity is determined by the depth of such blocks, $N_D$. Generally, a deeper neural network with a larger number of trainable parameters is able to learn a better generalization of the trained model, and consequently have a superior performance. On the other hand, an important question of the NN design is that, what is the minimal complexity that is necessary to achieve the desired accuracy. Therefore, two different complexities are investigated in this study: the Model~1 has $N_D=3$ depth and 1.13 million trainable parameters, while Model~2 consists of blocks with $N_D=5$ depth and 1.90 million trainable parameters.

Unlike in the original NN models and in conventional CV applications, where the main task is usually the classification of the input images, in our case the desired output is a set of physical quantities, observable in high-energy physics. Therefore, the activation function in the last layer of the models is the sigmoid function. The binary crossentropy is chosen for the loss function, while the Adam algorithm takes care for the optimization~\cite{kingma2017adam}. The models are implemented in Python, using Keras v2.4.0 with Tensorflow v.2.4.1 backend~\cite{chollet2015keras,abadi2016tensorflow}. The training, evaluating and testing were performed on a set of four Nvidia Tesla T4 graphics processing units of the Wigner Scientific Computational Laboratory (WSCLAB).

\section{Monte Carlo training data}
\label{sec:trainingdata}

Neither partonic degrees of freedoms, nor the hadronization process are directly (experimentally) observable; therefore, simulated e\-vents are needed to train and test the neural networks. For this purpose, the widely used {\sc Pythia} v8.3 general purpose event generator has been utilized with the commonly adopted Monash tune, that is known to reproduce LHC data with good accuracy~\cite{Sjostrand:1982fn, Sjostrand:2014zea, Skands:2014pea}. One of the main features of {\sc Pythia} is its string fragmentation model to perform the par\-ton-hadron transition~\cite{Sjostrand:1982fn}. The trained models are expected to generalize the main characteristics of this specific model. However, by future disentanglement of such a 'hadronizer' model might provide opportunities to study the unknown aspects of the hadronization process itself~\cite{Bellagente:2020piv}.

The simulated hadronic-state events were required to have at least two $R=0.6$ jets with $p_T\geq 40$~GeV/$c$ in the $|y|<(\pi - 0.4)$ region, following Ref.~\cite{Monk:2018zsb}, and as it is defined in the anti-$k_T$ algorithm~\cite{Cacciari:2008gp, Cacciari:2011ma}. Given this condition was satisfied, all final state particles were saved from the $y\in[-\pi, \pi]$, $\phi\in[0, 2\pi]$ region, both in the partonic and hadronic levels, just before and after the hadronization process.

For the training and validation dataset, proton-proton collisions with a center-of-mass energy $\sqrt{s}=7$ TeV were selected. For the predictions, $\sqrt{s}=0.9$~TeV,  $5.02$~TeV, and~$13$ TeV were considered -- at each collision energy and each individual dataset, 150\,000 events were generated with the conditions detailed above. Note that the total number of generated events, that did not satisfy the jet selection criteria, was larger with a factor of $\sim30$.

The parton-level events were discretized and normalized in the rapidity-azimuth-angle, $(\phi, y)$ plane to make them suitable for the CV algorithms: the $p=(E, p_x, p_y, p_z)$ four-momenta and the $m$ mass of the particles were summed in the given $(\Delta\phi, \Delta y)$ "pixel", with the multiplicity of that pixel as the 6\textsuperscript{th} "color channel" of the generated "pictures". The size of the image was chosen to be $31\times62$, resulting in a resolution of $\Delta\phi=0.2, \Delta y=0.1$. Due to the anisotropy in the azimuth angle, $\phi$, the half resolution in this direction is not expected to cause a significant effect. 

After the discretization process, each channel in each pixel was normalized into the $[0, 255]$ (continuous) region. On the hadron level, the event variables, described in Section~\ref{sec:jets}, were collected and scaled into the $[0, 1]$ region. Other preprocessing steps (such as jet centering, jet grooming, augmentation with translations and flips in the azimuthal coordinate) are not considered in this phase of the study.

\section{Results}
\label{sec:results}

\begin{figure}[H]
\centering
\includegraphics[width=0.9\linewidth]{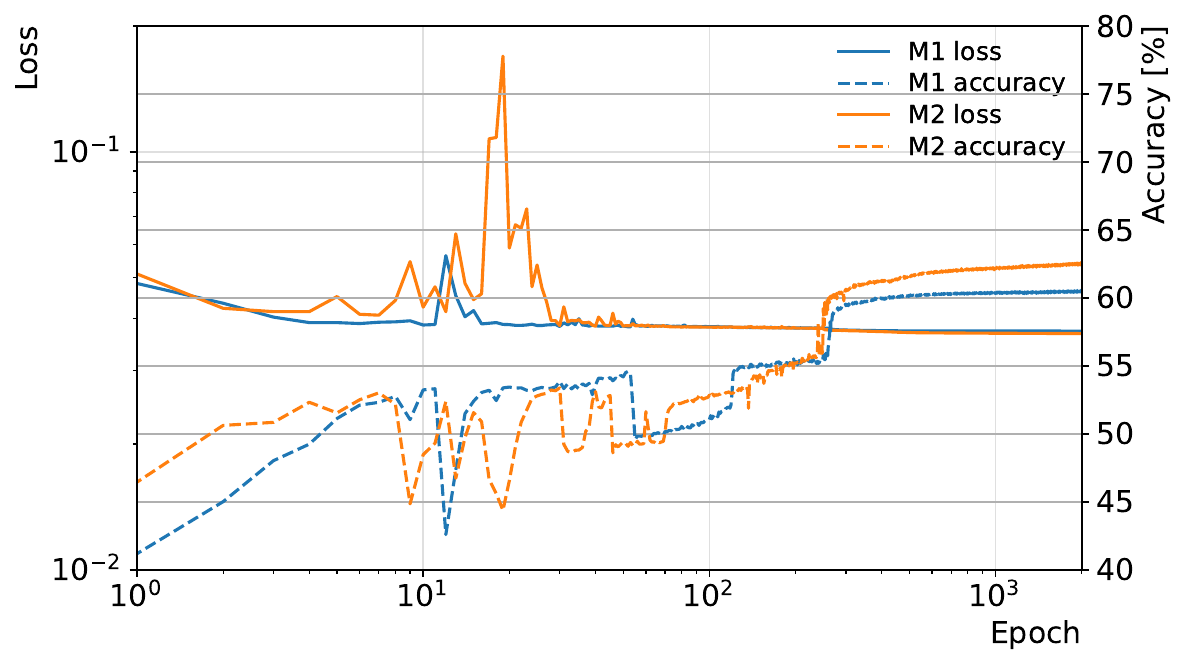}
\caption{The evolution of the loss and accuracy during the training epochs.}
\label{fig:losses}%
\end{figure}

The loss and accuracy values are shown on Fig.~\ref{fig:losses}. This logarithmic plot presents that, after $\sim$700 training epochs the loss (solid lines) and the accuracy (dashed lines) are saturating. This indicates that with the given NN complexity, further significant improvement can not be achieved within these nets, therefore the training of Model~1 and~2 were stopped after 2\,000 learning cycles. The trained models were validated on a different set of 150\,000 events at $\sqrt{s}=7$~TeV. 

\subsection{Global physical observables}
\label{subsec:mult}

\begin{figure}[H]
  \centering
  \includegraphics[width=\linewidth]{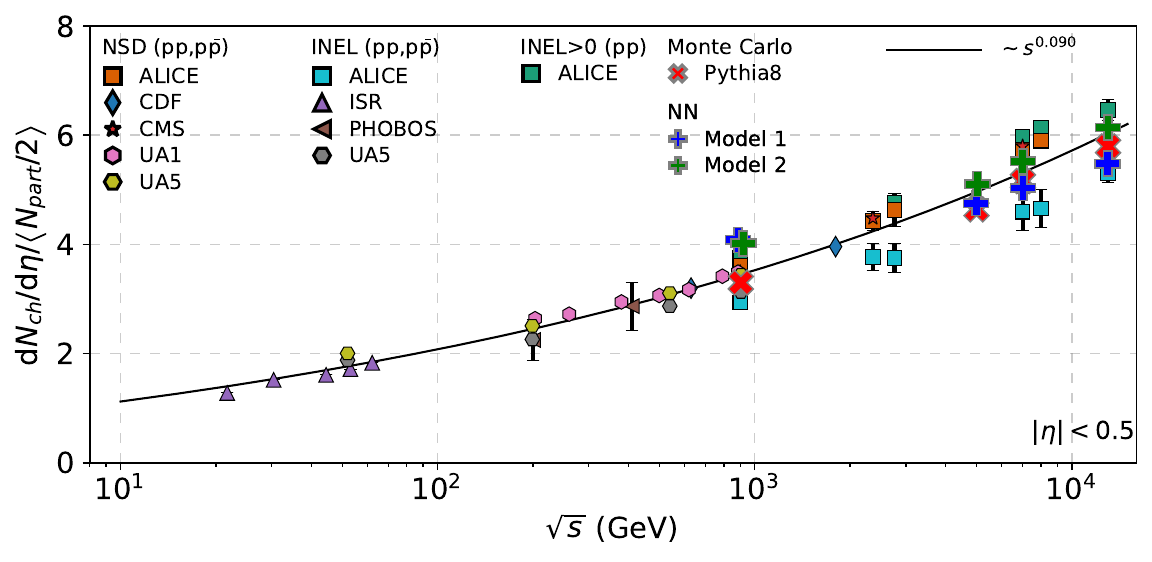}
  \caption{Charged particle density in the function of the center-of-mass energy form the collection in Ref.~\cite{ALICE:2012xs, ALICE:2015qqj}, {\sc Pythia8} Monte Carlo simulations, Model~1 and~2, respectively with red crosses, blue, and green daggers.}
  \label{fig:midrap}%
  \end{figure}

Recent studies showed that the average event multiplicity and its scaling is a well-observable fundamental quantity in high-energy collisions, which can be described analytically in the non-extensive statistical framework~\cite{Biro:2020kve, Vertesi:2020utz}. Various experimental measurements are also indicating that the char\-ged particle pseudorapidity density is increasing with the center-of-mass energy as 
\begin{equation}
\dd N_{ch}/\dd\eta\propto s_{NN}^a \ , 
\end{equation}
with $a\sim0.09-0.15$~\cite{ALICE:2012xs, ALICE:2015qqj} in the mid-rapidity, $|y|<0.5$ region. On Fig.~\ref{fig:midrap} the experimental data points of $\dd N_{ch}/\dd\eta/\langle N_{part}/2\rangle$ are plotted together with the Monte Carlo generated results (red crosses) and with the values predicted by the NNs (blue and green daggers). The predictions at collision energies, $\sqrt{s}=0.9$~TeV,  $5.02$~TeV, and~$13$ TeV, where Model~1 and~2 were not trained, show that both models were able to predict the hadron level event multiplicity within 10\% accuracy at LHC energies. 

The obtained Model~2 data points (green daggers) fit the $s_{NN}^{0.09}$ curve better, especially at the highest c.m. energies. As presented the structure of Model~2 on Fig.~\ref{fig:layout}(right), this network has more complexity than Model~1, therefore Model~2 presents better the logarithmic $\sqrt{s}$-scaling trend obtained by the Monte Carlo simulations and the data. 
One may conclude, that the complexity of the network Model~2 is getting close to represent the non-extensive evolution and the non-linear scaling properties of the hadronization process.

\subsection{Event shape and multiplicity distributions}
\label{eventshape}

While data on Fig.~\ref{fig:midrap} was restricted to the mid-rapidity region, more detailed information can be obtained on the event structure by investigating the event shape and multiplicity distributions in a wider $|y|<\pi$ range. By opening the rapidity area of interest, the low-multiplicity events also contribute to the event; indeed, a more valid picture can be obtained.

\begin{figure*}[thbp]
\centering

\includegraphics[width=0.24\linewidth]{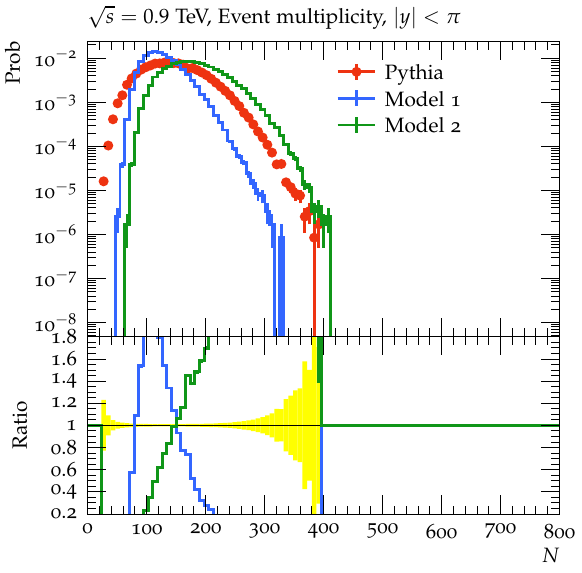}
\includegraphics[width=0.24\linewidth]{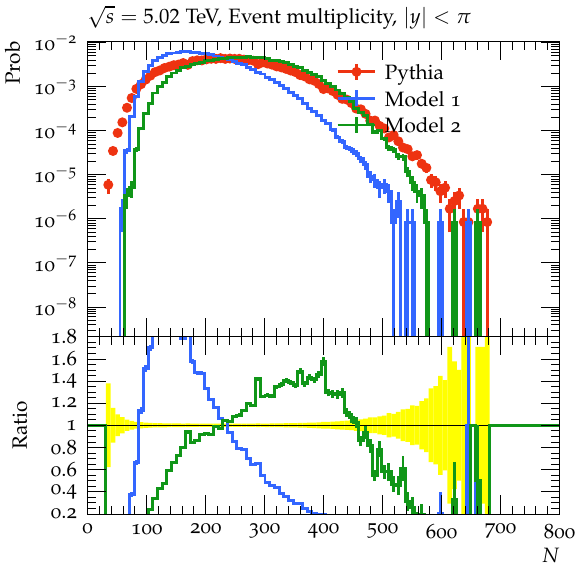}
\includegraphics[width=0.24\linewidth]{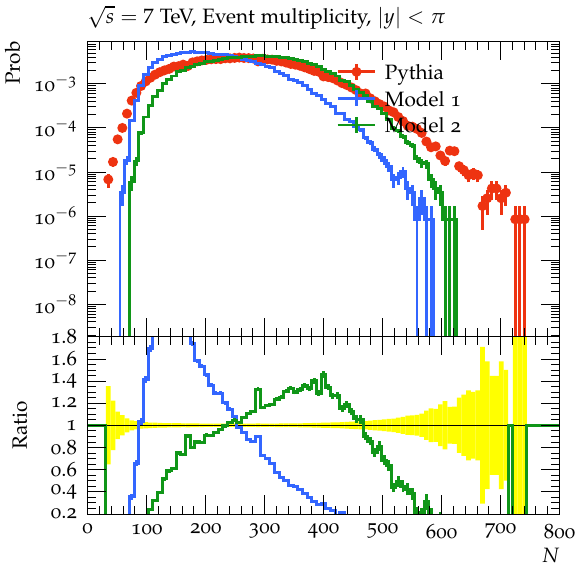}
\includegraphics[width=0.24\linewidth]{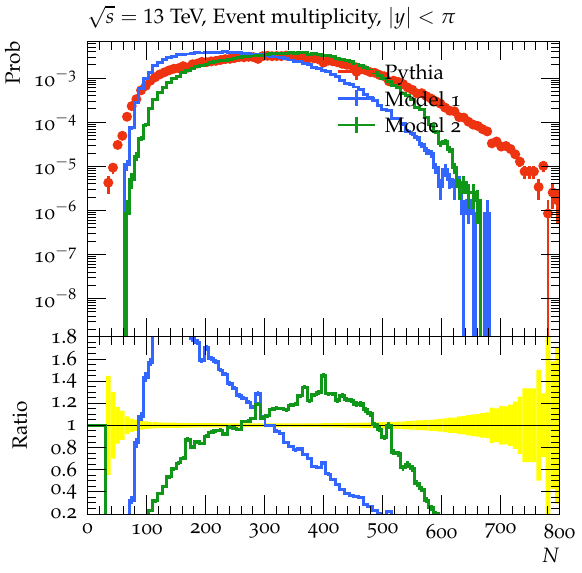}

\includegraphics[width=0.24\linewidth]{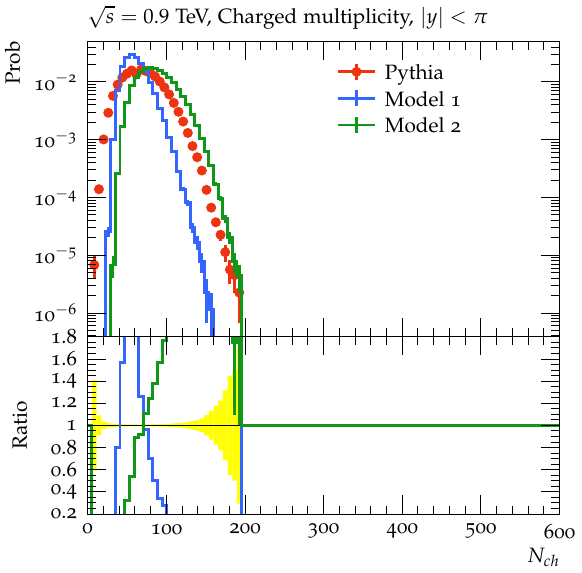}
\includegraphics[width=0.24\linewidth]{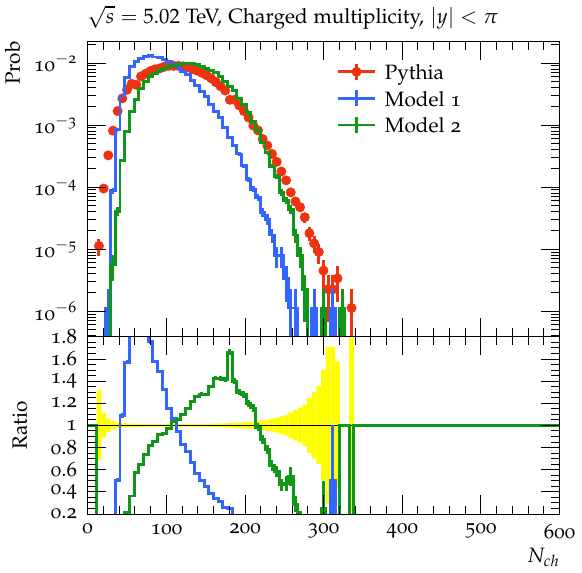}
\includegraphics[width=0.24\linewidth]{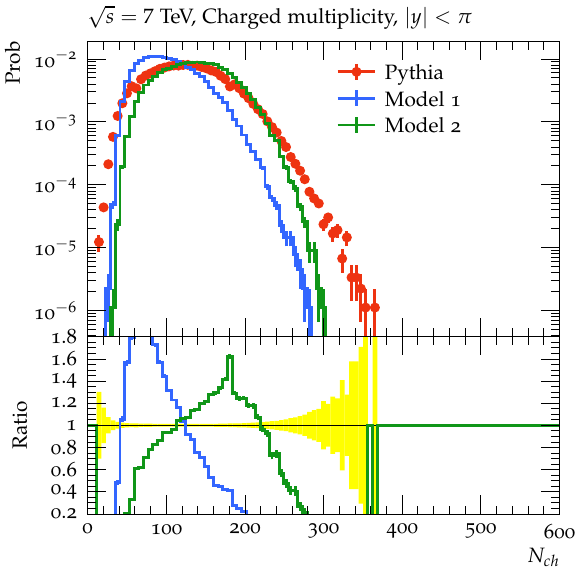}
\includegraphics[width=0.24\linewidth]{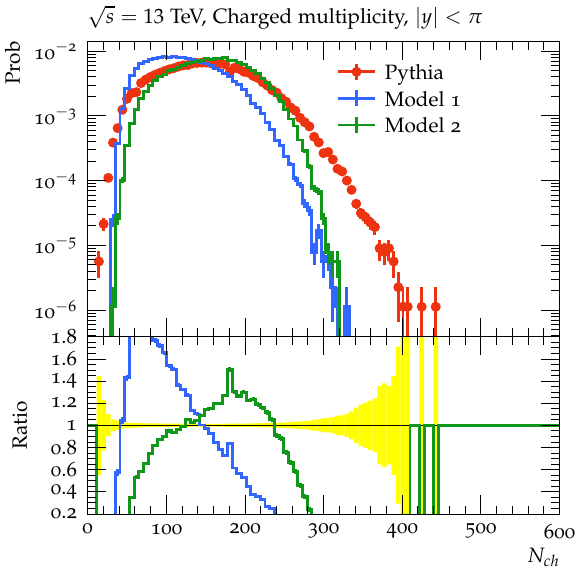}
\includegraphics[width=0.24\linewidth]{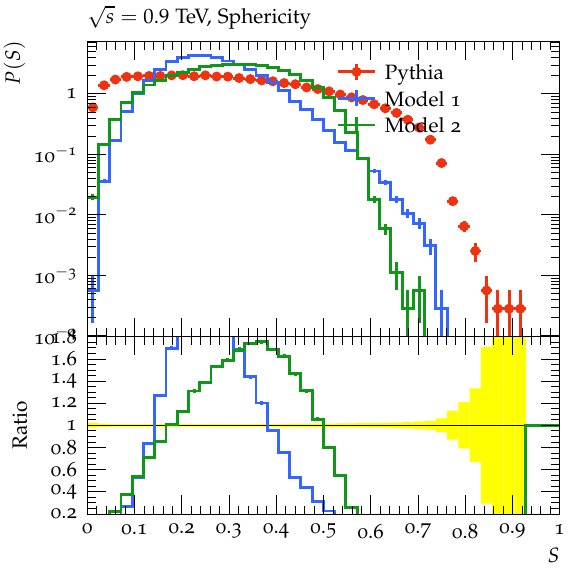}
\includegraphics[width=0.24\linewidth]{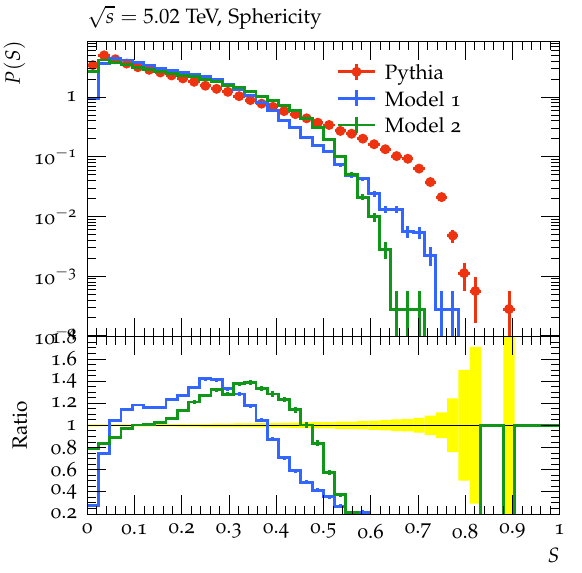}
\includegraphics[width=0.24\linewidth]{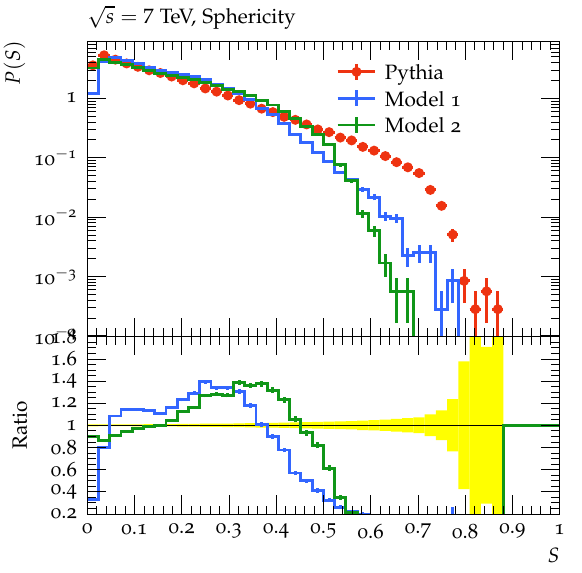}
\includegraphics[width=0.24\linewidth]{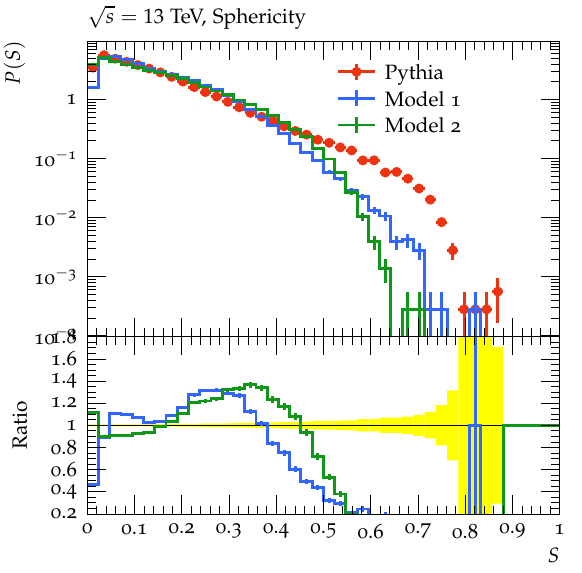}

\includegraphics[width=0.24\linewidth]{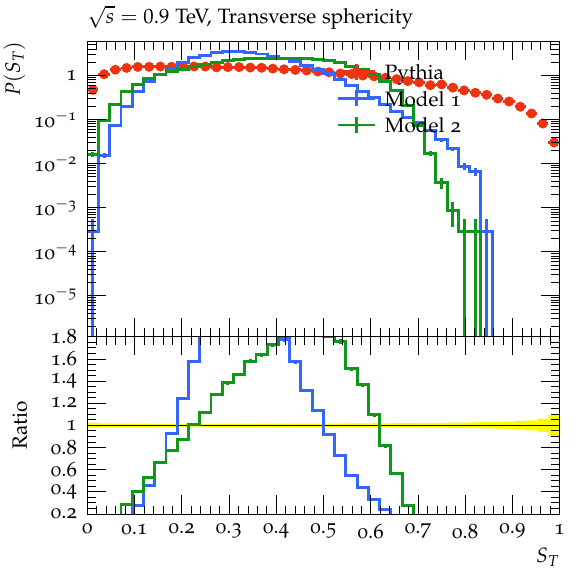}
\includegraphics[width=0.24\linewidth]{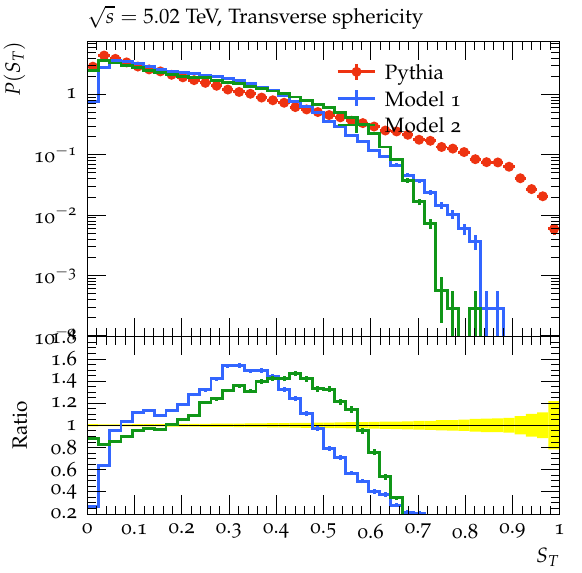}
\includegraphics[width=0.24\linewidth]{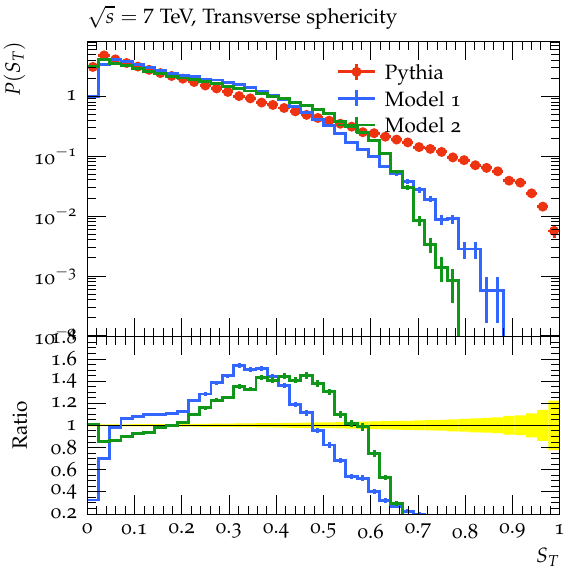}
\includegraphics[width=0.24\linewidth]{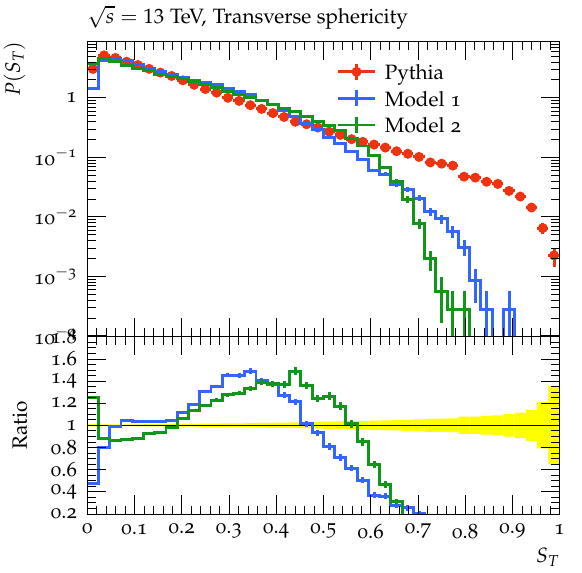}

  \caption{The neural network predicted multiplicity distributions and event shape variables compared with the reference Monte Carlo calculations in proton-proton collisions at $\sqrt{s}=0.9$ TeV (first column), $5.02$ TeV (second column), $7$ TeV (third column) and 13 TeV (fourth column).}
\label{fig:envvar1}%
\end{figure*}

On the panels of Fig.~\ref{fig:envvar1} Monte-Carlo-based simulations (red dots), ML-predicted multiplicity and event shape variable distributions are presented. We denote Model~1 and~2 using blue and green lines, respectively. Each row represents a distribution of a certain variable, and columns are for various collision energies in increasing order for $\sqrt{s}=0.9$~TeV, $5.02$~TeV, $7$~TeV, and~$13$ TeV. The third column presents the training energy, $\sqrt{s}=7$ TeV -- however, all plotted lines are based on the predictions from the trained neural network. Ratio panels show the fraction of the Model predictions relative to the Monte Carlo calculations on linear scale. On these Ratio panels yellow bands indicate the statistical uncertainty of the Monte Carlo data.

The event and (charged) particle multiplicity distributions are presented on the first two rows of the Fig.~\ref{fig:envvar1}. Statistical errors on these panels are increasing towards high multiplicities and end in the phase-phase cutoff. This is also true for the rare, low-multiplicity bins. Consequently, training will be more accurate and well-weighted in the intermediate multiplicity bins. This is represented well even in the training column at $\sqrt{s}=7$ TeV, where the predicted tails miss the Monte Carlo calculations, while the body of the distribution and the means are well predicted, especially for Model~2. Seemingly, by moving from the 1.13 million parameter Model~1 to the 1.9 million parameter Model~2, the ratios lie between $\pm 40\%$ within the well-trained multiplicity bins, $N_{ch},N\in [100,250]$. Overall for the Model~2, which has higher complexity, the expectation value of both the event- and the charged-particle multiplicity distributions are predicted within 10\% accuracy, similarly as it was for the rapidity-restricted case on Fig.~\ref{fig:midrap} as well. 

The NN-predictions for higher- and lower collision energies capture the qualitative properties of the total event multiplicity and charged multiplicity distributions. Since the lower part of the multiplicity distributions are pretty similar, the major deviation is rooted in the evolution of the tail influenced by the  phase-space and trainee-statistics cutoff. A KNO-like scaling can be observed on the NN-predictions as well, similarly as it was observed in jets~\cite{Vertesi:2020utz}. Thus, the network has adopted a non-linear scaling property.

The NN models successfully predict the event shape variable distribution on the last two rows of  Fig.~\ref{fig:envvar1}. Since it is hard to determine the highest sphericity at the sphericity limits, it is not surprising that best-agreed predictions are within the $\pm 40\%$ margin at the training energy, for bins $S,S_T\lesssim 0.5$. This is not unexpected though, since during the training only jetty events were considered, where at least 2 high energy jets were present in the hadron level event. The event-shape parameters present scaling for the collision-energy, however distributions at $\sqrt{s}=0.9$~TeV deviate well in the low- and high (transverse) sphericity values.

\subsection{Jet variables}

\begin{figure*}[thbp]
  \centering
  \includegraphics[width=0.24\linewidth]{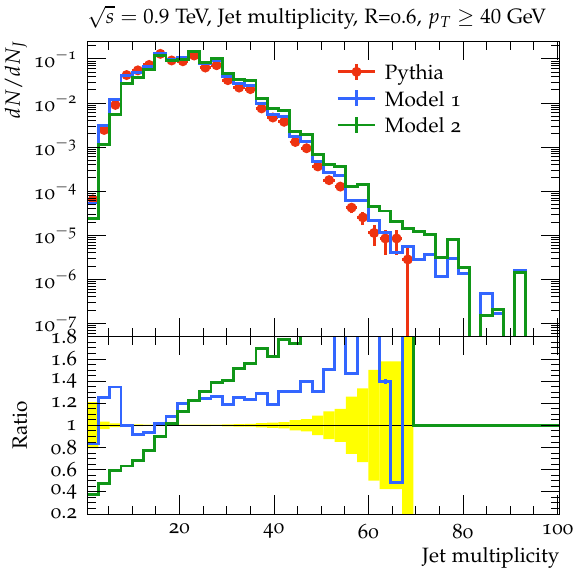}
  \includegraphics[width=0.24\linewidth]{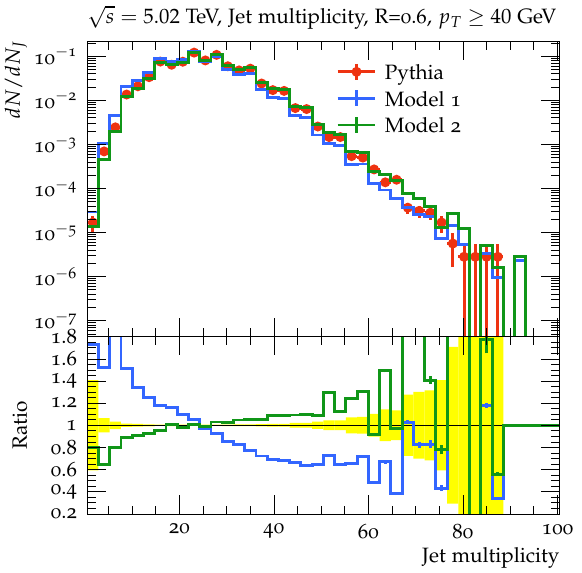}
  \includegraphics[width=0.24\linewidth]{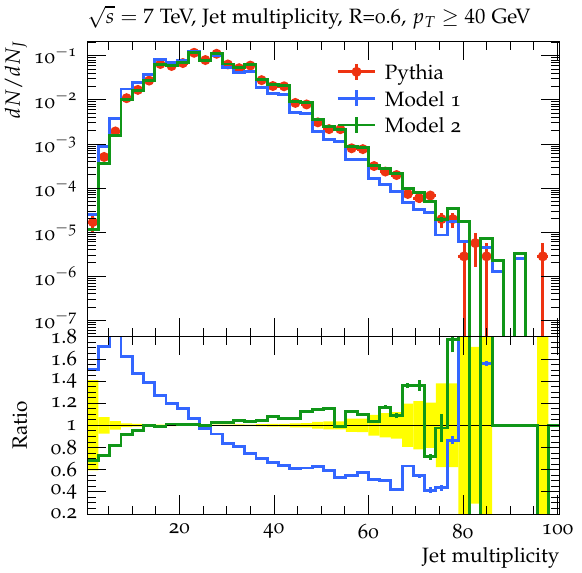}
  \includegraphics[width=0.24\linewidth]{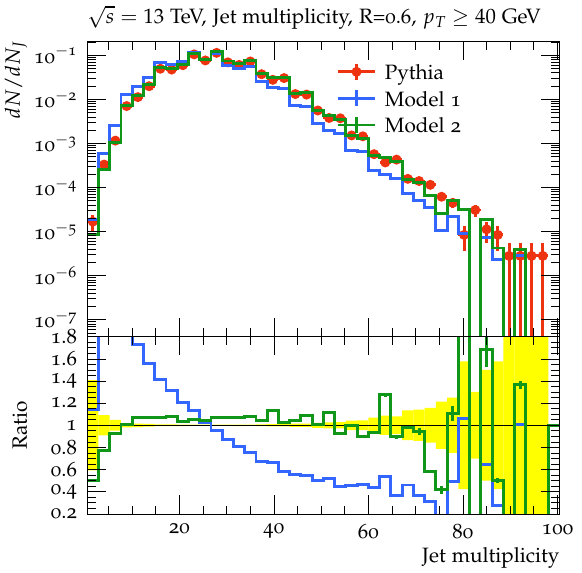}

  \includegraphics[width=0.24\linewidth]{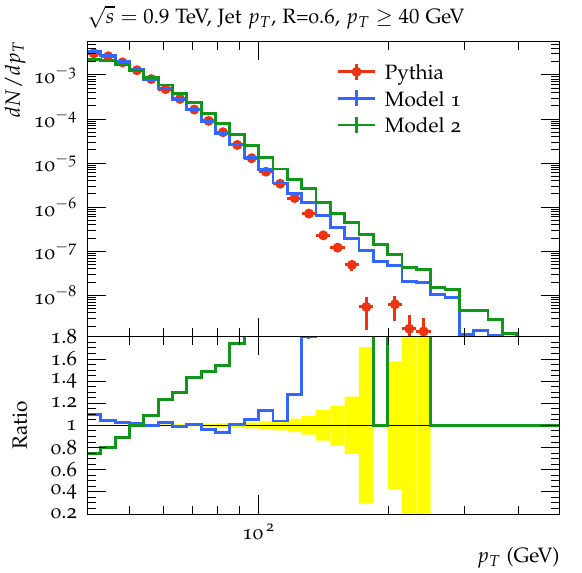}
  \includegraphics[width=0.24\linewidth]{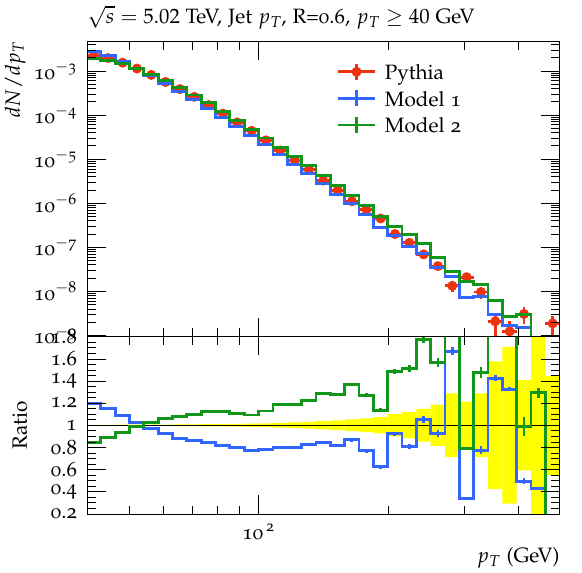}
  \includegraphics[width=0.24\linewidth]{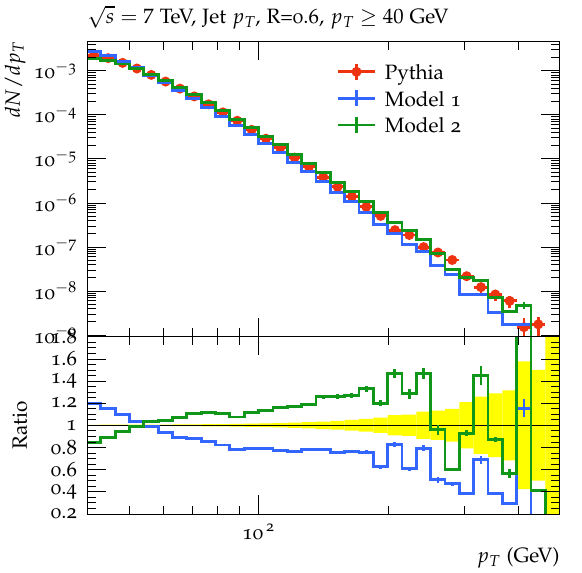}
  \includegraphics[width=0.24\linewidth]{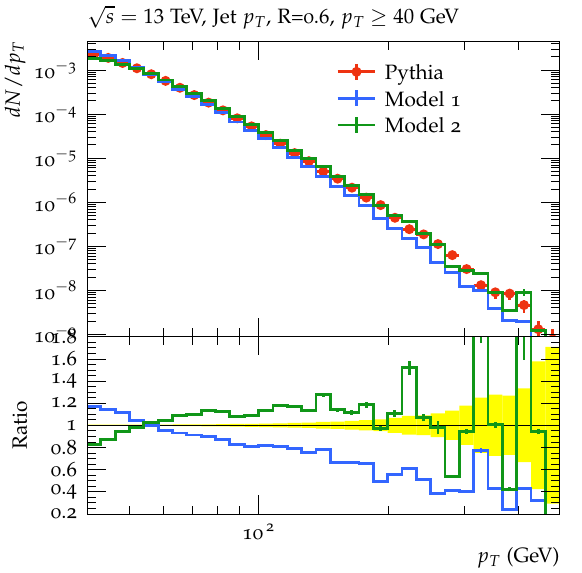}
  
  \includegraphics[width=0.24\linewidth]{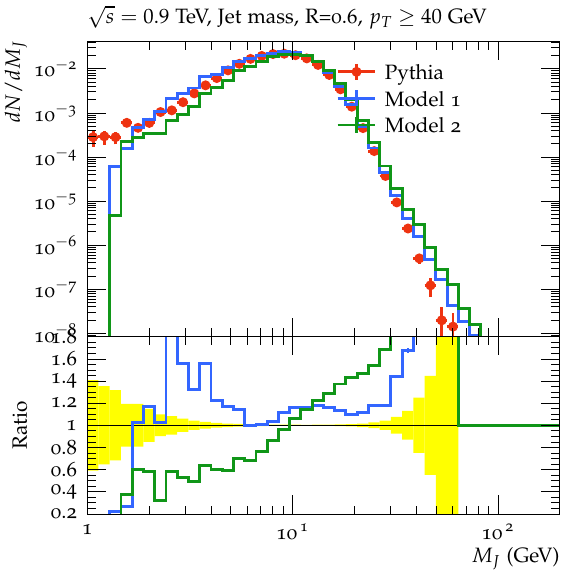}
  \includegraphics[width=0.24\linewidth]{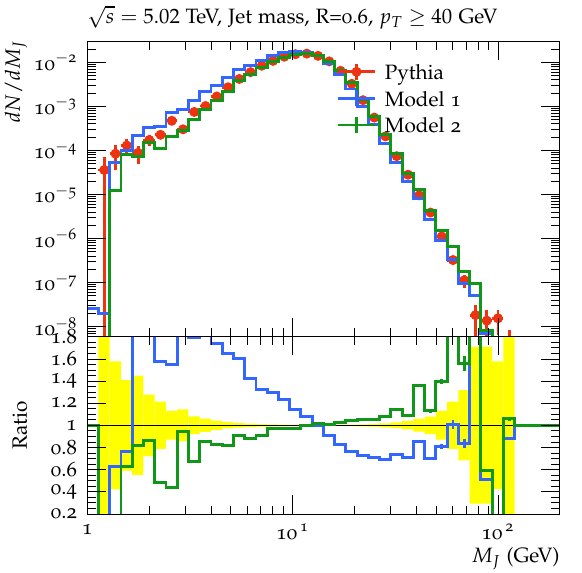}
  \includegraphics[width=0.24\linewidth]{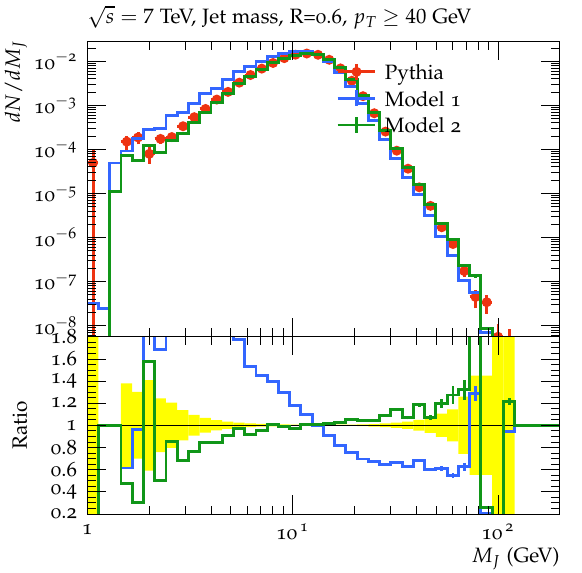}
  \includegraphics[width=0.24\linewidth]{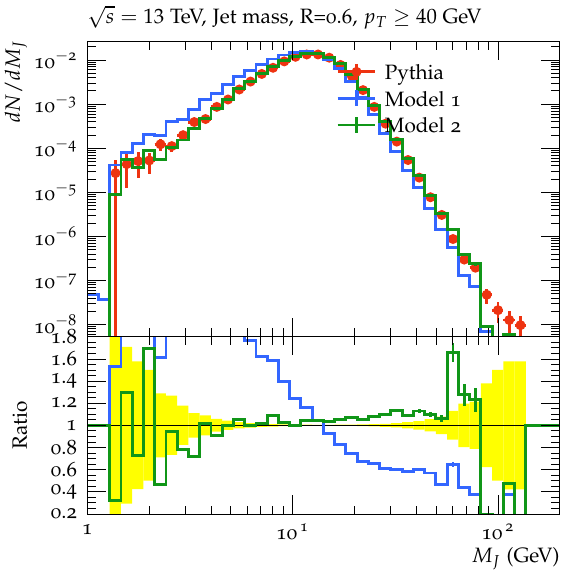}
  
  \includegraphics[width=0.24\linewidth]{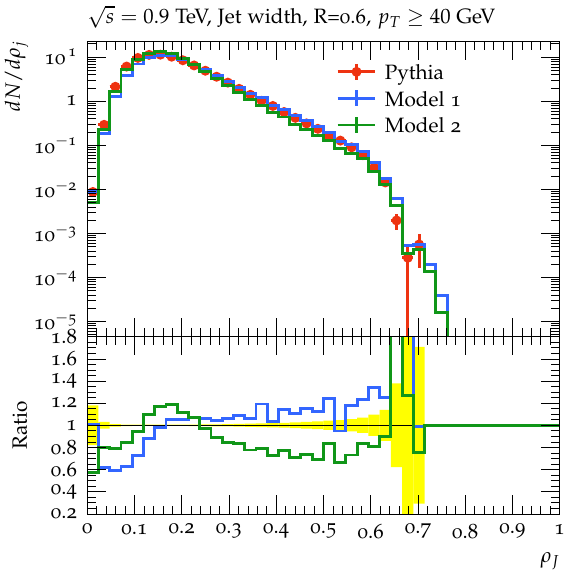}
  \includegraphics[width=0.24\linewidth]{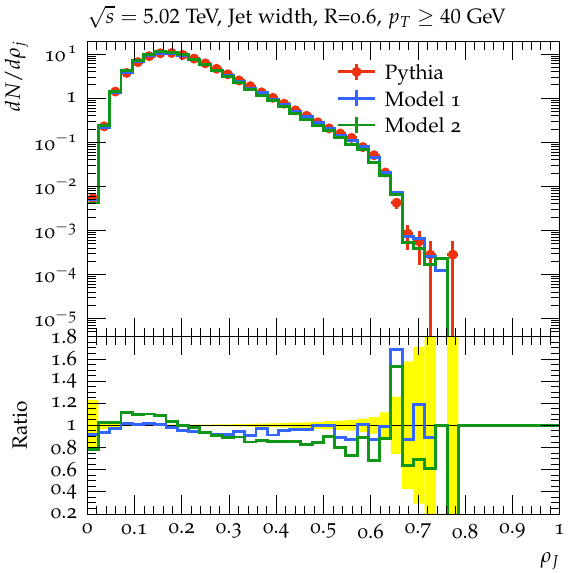}
  \includegraphics[width=0.24\linewidth]{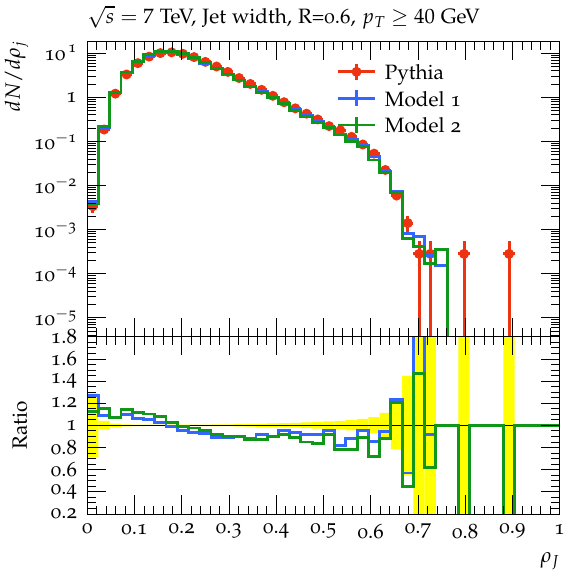}
  \includegraphics[width=0.24\linewidth]{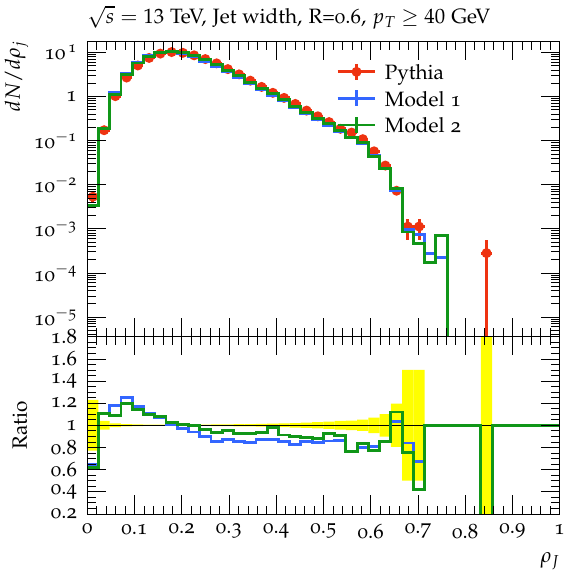}

  \caption{The neural-network predicted jet variables, compared with the reference Monte Carlo calculations at $\sqrt{s}=0.9$ TeV (first column), $5.02$ TeV (second column), $7$ TeV (third column) and 13 TeV (fourth column).}
  \label{fig:jetvar1}%
\end{figure*}

The jet variables, calculated by the reference Monte Carlo, are indicated on the panels of Fig.~\ref{fig:jetvar1}, together with the NN predictions. The outline of the panels and the markers are the same as in the previous section. Each row presents a certain jet-variable, for proton-proton collisions at c.m. energies from left to right, $\sqrt{s}=0.9$ TeV, $5.02$ TeV, $7$ TeV, and 13 TeV respectively. Rows are for: jet-multiplicity, jet mass, jet width and the transverse momentum distributions.

In general, the tested Models achieved both good qualitative and quantitative agreement for the investigated jet variables, calculated by the Monte Carlo simulations. At higher LHC energies, the accuracy is within 10\% for both Model~2 and $\sim$30\% for Model~1. The better agreement between the Monte Carlo and the NN is much more remarkable than in the level of global variables; indeed, it is true for all the investigated energies. The largest differences still follow the low-statistics part of the phase space indicated by yellow band on the panels. Since our whole study is focused on the hadronization and strongly connected with the jets, this precise matching is not surprising: input and output restricts well the jet properties in a limited phase-space. At the level of global observables, the space-phase is larger, therefore less constraint restricts the jet-trained networks.  

For the lowest energy, $\sqrt{s}=0.9$ TeV, the jet multiplicity (first row) and jet transverse momentum (second row) are slightly overestimated, while the jet mass and the jet width (third and fourth rows) still have good agreement. This indicates the importance of the jet substructure and underlying event studies~\cite{Mishra:2021hnr}. Moreover, the understanding of modifications in the jet structures is a key component of heavy-ion physics, where the change in the shape of the jet $p_T$ is traditionally understood as the sign of the onset of nuclear effects, therefore a trained solid baseline network would be appreciated well for further investigations.

At the higher energies, within the statistical uncertainties, Model~2 has a better agreement of jet mass with the reference Monte Carlo data. In contrast to that, the smaller-architecture Model~1 slightly overestimates the $M_J\lesssim10$ GeV region and underestimates above. The characteristics of the jet mass, together with the jet width, jet multiplicity and jet $p_T$ distributions are the main ingredients of jet classification methods. Energetic jets are typical hard probes of the quark-gluon plasma, therefore knowing the type of the jet -- whether is it originating from a quark or from a gluon -- is important to understand the nature of the medium interactions. Consequently, the investigated NN models might have such future applications too. 

Finally, at this level we can conclude that the presented networks of Model~1 and~2 are capable to learn hadronization patterns and scaling, especially at the jet level. In addition, the encoded parton-to-hadron representation can be applied well at the level of global variables if we apply to the high-statistics phase-space regions.

\section{Summary}
\label{sec:summary}

In this study, popular convolutional neural network architectures, widely used in Computer Vision applications, were investigated in the context of hadronization of partonic states in high-energy hadron collisions in two setups, with different complexities. The training and validation of the models were performed using simulated events by the {\sc Pythia 8} Monte Carlo event generator.

The simple models studied in this paper were able to learn the main features of the Lund string fragmentation. Moreover, the same trained models have been applied successfully to other center-of-mass energies, implying that the models could generalize the concept of multiplicity and energy scaling. It was also observed that learning of hadronization works on the level of global variables if the network sample the high-statistics part of the phase-space. This is a powerful feature, that can lead to various further research directions: a neural network with strong generalization capabilities could provide valuable input for the development and tuning process of Monte Carlo event generators. On the other hand, by reverse engineering the network, the properties of the process of hadronization might be investigated. In case of high-energy heavy-ion collisions, the same models might provide useful theoretical assist in the study of the parton-medium interactions as well.

\section*{Acknowledgement}
The research was supported by the Hungarian National Research, Development and Innovation Office (NKFIH) under the contract numbers OTKA K135515, K123815 and NKFIH 2019-2.1.11-T\'ET-2019-00078, 2019-2.1.11-T\'ET-2019-00050, the Wigner Scientific Computational Laboratory (WSCLAB) (the former Wigner GPU Laboratory). Author G. B. was supported by the Ministry of Innovation and Technology NRDI Office within the framework of the MILAB Artificial Intelligence National Laboratory Program. The authors are grateful to Gábor Papp and Antal Jakovác for the useful discussions.

\bibliography{papers_all}   

\end{document}